\documentclass[nonacm,sigplan]{acmart}

\usepackage[]{hyperref}
\usepackage{enumitem}
\usepackage{tabularx}
\usepackage{listings}
\usepackage{titlesec}
\usepackage{tikz}
\usepackage{makecell}
\usepackage{colortbl}
\usepackage{graphicx}
\usepackage{booktabs}
\usepackage{etoolbox}

\usepackage{subcaption}

\usepackage{xspace}

\definecolor{verylightgray}{rgb}{0.95,0.95,0.95}
\titleformat{\paragraph}[runin]{\normalfont\normalsize\bfseries}{\theparagraph}{0em}{}
\titlespacing*{\paragraph}{0pt}{0.25\baselineskip}{1em}

\setlist[itemize]{topsep=0.25\baselineskip, partopsep=0pt}
\setlist[enumerate]{topsep=0.25\baselineskip, partopsep=0pt}

\newcommand{\ecircled}[1]{\tikz[baseline=(char.base)]{\node[shape=circle,draw,inner sep=1pt,line width=0.8pt] (char) {#1};}}
\newcommand{\circled}[1]{%
  \tikz[baseline=(char.base)]{%
    \node[shape=circle,fill,inner sep=1pt] (char) {\textcolor{white}{#1}};%
  }%
}
\definecolor{lightgray}{gray}{0.9}
\newcommand{\lightmidrule}{\arrayrulecolor{lightgray}\midrule[\heavyrulewidth]\arrayrulecolor{black}}
\newcommand{\cmark}{\textcolor{green!70!black}{\scalebox{1.25}{$\boldsymbol{\checkmark}$}}}
\newcommand{\xmark}{\textcolor{red}{\scalebox{1.25}{$\boldsymbol{\times}$}}}

\newcommand{\graycell}{\cellcolor{lightgray}}

\newcommand{\tool}[0]{\textsc{Deep\allowbreak Context}\xspace}
\newcommand{\protocol}[0]{\mbox{\textsc{DLMonitor}}\xspace}

\begin{document}

\title{\tool: A Context-aware, Cross-platform, and Cross-framework Tool for Performance Profiling and Analysis of Deep Learning Workloads}


\author{Qidong Zhao}
\authornote{Both authors contributed equally to this research.}
\affiliation{%
  \institution{North Carolina State University}
  \city{Raleigh, NC}
  \country{United State}}
\email{qzhao24@ncsu.edu}

\author{Hao Wu}
\authornotemark[1]
\affiliation{%
  \institution{George Mason University}
  \city{Fairfax, VA}
  \country{United State}}
\email{hwu27@gmu.edu}

\author{Yueming Hao}
\affiliation{%
  \institution{North Carolina State University}
  \city{Raleigh, NC}
  \country{United State}}
\email{yhao24@ncsu.edu}

\author{Zilingfeng Ye}
\affiliation{%
  \institution{George Mason University}
  \city{Fairfax, VA}
  \country{United State}}
\email{yipzlf@gmail.com}

\author{Jiajia Li}
\affiliation{%
  \institution{North Carolina State University}
  \city{Raleigh, NC}
  \country{United State}}
\email{jiajia.li@ncsu.edu}

\author{Xu Liu}
\affiliation{%
  \institution{North Carolina State University}
  \city{Raleigh, NC}
  \country{United State}}
\email{xliu88@ncsu.edu}

\author{Keren Zhou}
\affiliation{%
  \institution{George Mason University}
  \city{Fairfax, VA}
  \country{United State}}
\email{kzhou6@gmu.edu}

\begin{abstract}
Effective performance profiling and analysis are essential for optimizing training and inference of deep learning models, especially given the growing complexity of heterogeneous computing environments.
However, existing tools often lack the capability to provide comprehensive program context information and performance optimization insights for sophisticated interactions between CPUs and GPUs.
This paper introduces \tool{}, a novel profiler that links program contexts across high-level Python code, deep learning frameworks, underlying libraries written in C/C++, as well as device code executed on GPUs.
\tool{} incorporates measurements of both coarse- and fine-grained performance metrics for major deep learning frameworks, such as PyTorch and JAX, and is compatible with GPUs from both Nvidia and AMD, as well as various CPU architectures, including x86 and ARM.
In addition, \tool{} integrates a novel GUI that allows users to quickly identify hotpots and an innovative automated performance analyzer that suggests users with potential optimizations based on performance metrics and program context.
Through detailed use cases, we demonstrate how \tool{} can help users identify and analyze performance issues to enable quick and effective optimization of deep learning workloads.
We believe \tool{} is a valuable tool for users seeking to optimize complex deep learning workflows across multiple compute environments.
\end{abstract}

\maketitle 
\pagestyle{plain} 

\section{Introduction}
\label{sec:introduction}
The rapid advancement of deep learning has led to increasingly complex models~\cite{achiam2023gpt,dubey2024llama,peebles2023scalable} deployed across diverse and heterogeneous computing environments.
Optimizing the training and inference of these models is critical for improving performance and reducing computational costs~\cite{hoffmann2022training,kaplan2020scaling}.
However, the sophisticated interactions between CPUs and GPUs, coupled with the diversity of frameworks~\cite{jax2018github,paszke2019pytorch} and compilation modes~\cite{ansel2024pytorch}, pose significant challenges for developers seeking to identify and address performance bottlenecks effectively.

To improve the efficiency of deep learning workloads by fully utilizing hardware resources, effective performance profiling tools are essential.
These tools include framework-specific solutions, such as the PyTorch profiler~\cite{pytorch_profiler} and the JAX profiler~\cite{jax_profiler}, as well as those provided by hardware vendors, like Nsight Systems~\cite{nsight_systems}, Roctracer~\cite{roctracer}, and VTune~\cite{vtune}.
The primary functionality of these tools is tracing, which captures metrics associated with individual CPU and GPU operations and displays them on a comprehensive timeline to assist users in investigating performance bottlenecks.
\begin{table*}[t]
    \centering
    \footnotesize
    \caption{Comparison of \tool{} (our tool) with existing profiling tools.}
    \begin{tabularx}{\textwidth}{X *{7}{>{\centering\arraybackslash}X}}
    \toprule
        \makecell{\textbf{Profiling}\\\textbf{Tools}}
        & \makecell{\textbf{Python}\\\textbf{Context}}
        & \makecell{\textbf{Framework}\\\textbf{Context}}
        & \makecell{\textbf{C++}\\\textbf{Context}}
        & \makecell{\textbf{Device}\\\textbf{Context}}
        & \makecell{\textbf{Cross}\\\textbf{GPUs}}
        & \makecell{\textbf{Cross}\\\textbf{Frameworks}}
        & \makecell{\textbf{CPU}\\\textbf{Profiling}} \\
    \midrule
        \mbox{\textbf{Nsight Systems~\cite{nsight_systems}}}       & \cmark & \xmark & \cmark & \xmark & \xmark & \cmark & \cmark \\
    \lightmidrule
        \mbox{\textbf{RocTracer~\cite{roctracer}}}  & \xmark & \xmark & \xmark & \xmark & \xmark & \xmark & \xmark \\
    \lightmidrule
        \mbox{\textbf{JAX profiler~\cite{jax_profiler}}}        & \cmark & \xmark & \xmark & \xmark & \cmark & \xmark & \cmark \\
    \lightmidrule
        \mbox{\textbf{PyTorch profiler~\cite{pytorch_profiler}}}    & \cmark & \cmark & \xmark & \xmark & \cmark & \xmark & \cmark \\
    \lightmidrule
        \graycell\mbox{\textbf{\tool{}}} & \graycell\cmark & \graycell\cmark & \graycell\cmark & \graycell\cmark & \graycell\cmark & \graycell\cmark & \graycell\cmark \\
    \bottomrule
    \end{tabularx}
    \label{tab:tool_comparison}
\end{table*}

\begin{figure}[t]
    \centering
    \begin{subfigure}[h]{\linewidth}
        \includegraphics[width=\linewidth]{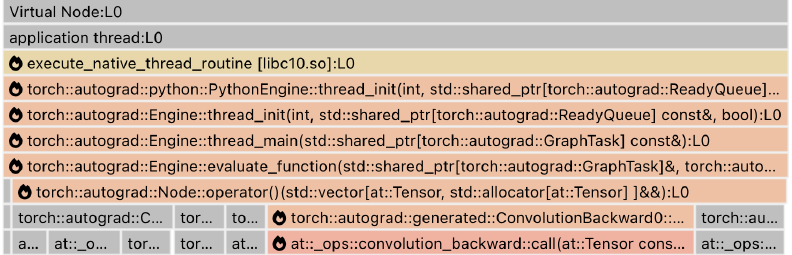}
        \caption{The hot call path w/o framework context}
        \label{fig:wo_torch_monitor_callpath}
    \end{subfigure}
    \begin{subfigure}[h]{\linewidth}
        \includegraphics[width=\linewidth]{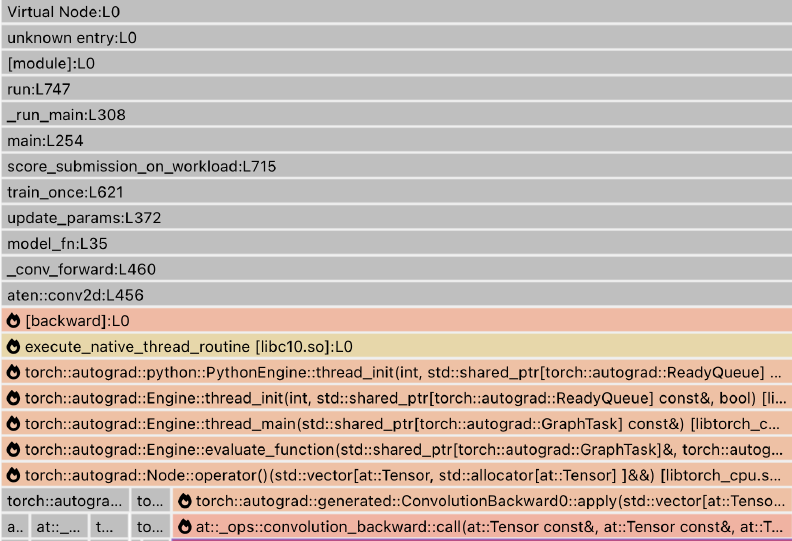}
        \caption{The hot call path w/ framework context}
        \label{fig:w_torch_monitor_callpath}
    \end{subfigure}
    \caption{Comparison of highlighted call paths w/ and w/o framework context. The thicker the color of a frame, the more time has been spent on that frame.}
    \label{fig:callpath-comparison}
\end{figure}

The design of existing tools presents several challenges that hinder thorough performance analysis and optimization.
We have summarized their features and limitations in Table~\ref{tab:tool_comparison}.

First, they often fail to provide a comprehensive view of performance metrics across the entire software stack, which spans from high-level Python code and frameworks down to low-level C++ operations and GPU instructions.
At best, Nsight Systems can correlate Python with C++ code, but it cannot provide information about deep learning operators and GPU instructions.
Figure~\ref{fig:callpath-comparison} illustrates this limitation.
In Figure~\ref{fig:wo_torch_monitor_callpath}, only C++ code is visible in the call path, and it is unclear where the convolution function was called without framework information.
This is because the backward and forward operators are launched from different CPU threads in PyTorch.
Such a limitation obscures the pinpointing of performance problems related to specific convolution operations, especially when a complex model may invoke hundreds of convolution operations~\cite{he2016deep}.
In contrast, Figure~\ref{fig:w_torch_monitor_callpath} reveals the Python call path and associated deep learning operators, providing deeper insights for performance optimization.

Second, most existing performance tools trace and record every operation, causing huge profiles when profiling long-running training workloads.
Identifying issues within traces that contain millions of operations often requires a great deal of manual efforts.
Moreover, the voluminous profile data can consume significant memory resources, potentially exhausting DRAM capacity or making it impossible for visualization.
Existing tools can only aggregate metrics (e.g., time) postmortem for individual kernels, but they fail to streamline metrics aggregation online to reduce profile size and do not differentiate between instances called from different contexts for automated problem identification.

Lastly, existing tools lack portability across platforms and frameworks.
Vendor-provided tools, such as Nvidia's Nsight suite, can profile events from JAX and PyTorch on Nvidia GPUs, but they are incompatible with workloads running on AMD GPUs.
On the other hand, framework-specific tools like the PyTorch profiler can support both AMD and Nvidia GPUs but cannot handle workloads written in JAX.
This fragmentation in tool availability increases the learning curve for users and limits their ability to cross-reference findings across different frameworks and GPUs, making it difficult to determine which setup best suits their workloads.

 



\sloppy
In response to these challenges, we introduce \tool{}, a novel profiler that delivers comprehensive performance insights for deep learning workloads.
Unlike existing profilers that lack essential context, \tool{} captures all critical program context information relevant to deep learning workloads.
As shown in Table~\ref{tab:tool_comparison}, \tool{} enables the identification of performance issues in multiple layers of the software stack, spanning high-level Python code, deep learning frameworks, underlying libraries written in C/C++, and device code executed on GPUs.
Additionally, \tool{} supports performance profiling for major deep learning frameworks including PyTorch and JAX, AMD and Nvidia GPUs, as well as x86 and ARM CPUs. 
Finally, \tool{} enables the collection, attribution, and aggregation of both coarse and fine-grained performance metrics, allowing for detailed investigation of performance bottlenecks with minimal profile data, even for long-running applications.

This paper presents the design, implementation, and evaluation of \tool{} and makes the following research contributions:

\begin{itemize}[leftmargin=\parindent]
    \item We introduce a ``shim'' layer---\protocol{}---that converts deep learning framework-specific data into a framework-agnostic format, enabling seamless integration of framework information with third-party performance tools. 
    \item We design an automated performance analyzer that provides actionable optimization suggestions based on performance metrics and program contexts, such as fusing operators, changing data layouts, or modifying hardware configurations.
    \item We describe a novel graphical user interface (GUI) that visualizes performance data in a compact, navigable, and hierarchical format, allowing users to quickly identify performance bottlenecks and apply optimizations based on performance analysis results.
\end{itemize}

We evaluated \tool{} using a diverse range of deep learning workloads across various platforms and frameworks, demonstrating that \tool{} significantly saves memory and disk space usage with similar runtime overhead compared to the state-of-the-art performance tools.
Through use case studies, we show that \tool{} can effectively identify performance issues and suggest insightful code changes, enabling straightforward optimization of deep learning models.
Even users with limited experiences in deep learning frameworks or CPU and GPU architectures can achieve speedups ranging from 1.06$\times$ to 1.66$\times$.

\section{Related Work}
\label{sec:related work}
In this section, we review existing approaches about deep learning profilers, call path profiling, and automated performance analysis.

\paragraph{Deep Learning Profilers}
Many profilers can measure deep learning workloads. These tools include vendor-provided tools such as Nsight Systems~\cite{nsight_systems}, VTune~\cite{vtune}, RocTracer~\cite{roctracer} and DLProf~\cite{nvidia_dlprof}, as well as framework-based tools such as the JAX profiler~\cite{jax_profiler} and the PyTorch profiler~\cite{pytorch_profiler}.
However, these tools often focus on specific frameworks or platforms with limited applicability.

Some workload-specific profilers, such as RL-Scope~\cite{gleeson2021rl} and XSP~\cite{li2020xsp}, analyze interactions across layers of the deep learning stack to identify bottlenecks that are not obvious when examining individual layers.
\tool{} advances these tools by employing a generic solution for different frameworks, GPUs, and platforms.


There are profilers that post-process metrics from existing profiling tools.
For instance, Hotline Profiler~\cite{snider2023hotline} introduces a multi-scale timeline with annotations for DNN training, based on the postmortem analysis of results from the PyTorch profiler~\cite{pytorch_profiler}.
Similarly, DLProf~\cite{nvidia_dlprof} analyzes the results collected from Nsight Systems.
Since these tools do not modify the runtime, they suffer from the same limitations as trace-based profilers, which incur significant memory and disk overhead.

Skyline~\cite{yu2020skyline}, as the most related tool to our tool, offers an interactive profiling experience by integrating the profiler into the development environment.
However, unlike \tool{} that intercepts ``native'' C/C++ operations, Skyline uses monkey patching~\cite{} for PyTorch Python operations, which introduces overhead and prevents it from obtaining native call path information.
Additionally, it does not interact with vendor-provided profiling substrates, limiting its ability to gather abundant low-level information using performance counters.

\paragraph{Call Path Profilers}
General performance tools such as HPCToolkit~\cite{zhou2021measurement}, TAU~\cite{shende2006tau}, perf~\cite{linux_perf}, and DrCCTProf~\cite{zhao2020drcctprof} offer call path profiling and performance analysis for low-level languages such as Fortran and C/C++.
These tools provide deep insights into the complex behaviors of the underlying software stack and operating system, offering a complementary perspective to what deep learning profilers may miss.
Additionally, some of these tools can sample a large set of CPU performance counters, going beyond coarse-grained metrics.
However, these tools often lack integration with Python runtime and deep learning frameworks, limiting their effectiveness in profiling multi-language environments.

On the other hand, Python-specific profiling tools like Scalene~\cite{berger2020scalene} and cProfile~\cite{python_cprofile} provide effective analysis of Python call paths but lack the ability to analyze call paths in lower-level languages.
Moreover, they can only collect limited information about accelerators.
\tool{} addresses these shortcomings by providing a comprehensive call context that spans every level of the software stack, effectively bridging the gap between native language profiling and high-level language analysis.

\paragraph{Automated Performance Analysis}

Existing automated performance analysis tools often target specific or limited domains.
For instance, tools such as Nsight Compute~\cite{nvidia_nsight_compute} and GPA~\cite{zhou2021gpa} focus on pattern matching of GPU kernels using expert-defined rules based on fine-grained metrics.
These tools provide insights into how and where to modify kernel source code to improve performance.
On the other hand, tools such as Nsight Systems~\cite{nsight_systems} and DLProf~\cite{nvidia_dlprof} analyze trace patterns to provide coarse-grained recommendations.
They offer insights into which GPU operations are expensive and whether the CPU is causing performance bottlenecks. However, these tools do not consider both low- and high-level contexts simultaneously.
In contrast, \tool{} introduces a pattern matching system that allows flexible rules to be defined for analysis, incorporating low- and high-level contexts based on fine- and coarse-grained metrics.
This approach provides a more holistic view of performance issues, enabling more flexible analysis and effective optimizations.


\section{Overview}
\label{sec:overview}

\tool{} is designed to achieve multiple objectives, including providing a holistic view that spans high- and low-level contexts, enabling cross-framework profiling, and supporting fast or even automated performance analysis.
Each of these objectives presents unique challenges: (1) High- and low-level contexts are obtained in different methods, as Python is interpreted while C++ is compiled, and framework-specific information cannot be obtained simply by examining either C/C++ or Python code alone.
(2) Different frameworks are implemented in vastly different ways, making direct instrumentation of their source code to capture information both unstable and unmaintainable.
(3) For complex deep learning workloads, the profiling database will include extensive program context, numerous GPU kernel invocations, and various metrics, making it challenging to manually identify performance issues.

\begin{figure}[ht]
    \centering
    \includegraphics[width=0.8\linewidth]{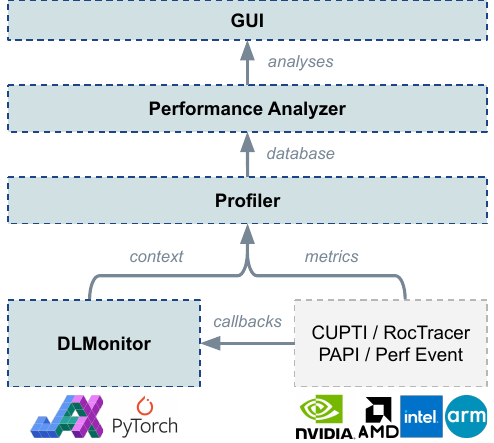}
    \caption{Major components of \tool{}}
    \label{fig:overview}
\end{figure}
To address these challenges, \tool{} is structured into four primary components, as shown in Figure~\ref{fig:overview}, each designed to handle distinct aspects of performance profiling for deep learning workloads.
The \textit{profiler} is responsible for collecting, attributing, and aggregating performance metrics.
It gathers performance metrics from Nvidia and AMD GPUs through the CUPTI~\cite{cupti} and RocTracer~\cite{roctracer} APIs, and from CPUs through Linux system calls, Perf events~\cite{linux_perf}, and the PAPI API~\cite{weaver2012measuring}.
Once the program context is obtained from \protocol{}, the profiler associates the metrics collected with the program context and aggregates the metrics within the same context when necessary.
\protocol{} serves as a ``shim'' layer between the profiler and the deep learning framework.
It allows the profiler to intercept framework operations during the execution of deep learning frameworks and provides a comprehensive context when the profiler calls \protocol{}.
Next, the \textit{performance analyzer} processes the collected data postmortem, identifying performance bottlenecks and optimization opportunities. Finally, the \textit{GUI} presents the analyzed data in an intuitive format, compatible with Visual Studio Code~\cite{vscode}, to facilitate interpretation and inspection of performance issues.

\section{Design and Implementation}
\label{sec:design and Implementation}
This section is organized into four subsections, each dedicated to one of the aforementioned modules of \tool. 

\subsection{\protocol{}}
\begin{figure}[ht]
    \centering
    \includegraphics[width=0.9\linewidth]{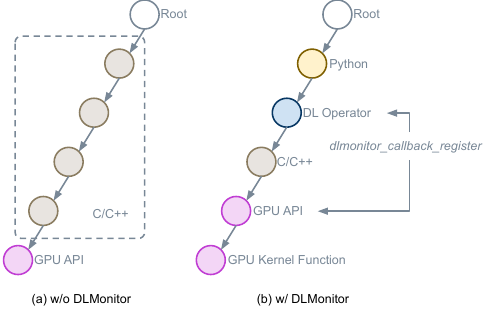}
    \caption{Comparison between call paths w/ and w/o \protocol{}}
    \label{fig:call path}
\end{figure}
\protocol{} is a key component of \tool{}, providing a unified interface for obtaining call paths and registering callbacks within deep learning frameworks.
Profilers interact with deep learning frameworks by invoking \protocol{}'s APIs.
The core APIs include:

\begin{itemize}[leftmargin=\parindent]
    \item \texttt{dlmonitor\_init}: Initializes \protocol{}'s shared library, \texttt{libdlmonitor.so}, which is typically loaded at the start of execution using utilities like \texttt{LD\_PRELOAD}.
    \item \texttt{dlmonitor\_callback\_register}: Registers the callback specified by profilers to intercept operations in a specific domain, such as deep learning frameworks and GPU runtime.
    \item \texttt{dlmonitor\_finalize}: Disables \protocol{} monitoring and releases all interceptions.
    \item \texttt{dlmonitor\_callpath\_get}: Constructs and returns a multi-layer call path to the profiler.
\end{itemize}

In Figure~\ref{fig:call path} (b), we show an example call path constructed by \protocol{}, which includes frames from Python, frameworks, C/C++, GPU kernels, and execution within GPU kernels.
Note that without \protocol{}, the call path we obtain in Figure~\ref{fig:call path} (a) only contains C/C++ frames (i.e., native call path), without information about Python, deep learning frameworks, as well as execution within GPU kernels.
wProfiling tools can register callbacks at individual framework operations and underlying GPU APIs to obtain the call path and gather necessary information for performance analysis.
In the following, we describe implementation details about how \protocol{} intercepts operations and constructs call paths.

\paragraph{Intercepting Framework Operations}
\protocol{} intercepts operations in PyTorch and JAX, allowing profilers to register callbacks using the \texttt{dlmonitor\_callback\_register} function before and after each operation.
Here, the the domain provided to the function is \texttt{DLMONITOR\_FRAMEWORK}.
Interception points include individual deep learning operators (e.g., \texttt{torch.matmul}), the start and end of compute graph compilation, and tensor memory allocation/deallocation.
At these points, profilers can access information such as operators inputs and outputs, and retrieve the full call path via the \texttt{dlmonitor\_callpath\_get} function.

\begin{figure}[ht]
    \centering
    \includegraphics[width=\linewidth]{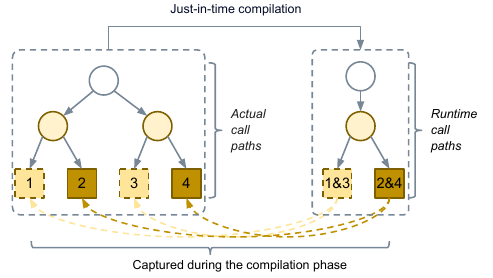}
    \caption{\protocol{} intercepts JAX's compilation phase to map fused operators to original ones.}
    \label{fig:jax_monitor}
\end{figure}

\protocol{} employs different mechanisms to support PyTorch and JAX, ensuring compatibility with frameworks installed via \texttt{pip} wheels \textit{without source code modifications}.
For PyTorch, it leverages PyTorch's \texttt{aten::addGlobalCallback} interface, which allows for invoking customizable callbacks at various points.
Unlike PyTorch, implementing \protocol{} for JAX presents two challenges.
First, JAX does not inherently support registering callbacks before and after each deep learning operator. 
Second, while PyTorch's eager mode is widely used, where each operator is executed individually, JAX compiles operators into computation graphs with fused operations before execution.
Once compiled, the runtime call path of each operator differs from its call path in the original code where it was compiled from.
To address these challenges, we implemented a lightweight binary instrumentation utility that allows JAX to provide profiling information comparable to that of PyTorch.
Specifically, to address the first challenge, we utilize binary instrumentation to intercept JAX's compilation function for passes of computation graphs and insert callbacks before and after each JAX operator after the very last pass.
For the second challenge, we record the mappings between fused operators to original ones (Figure~\ref{fig:jax_monitor}) in the operator fusion pass.
The call path of each original operator is recorded during the compilation phase, while the call path of the fused operator is recorded at runtime.
In the GUI, we display all possible original call paths associated with the runtime call path of each fused operator.

\paragraph{Intercepting GPU APIs}
In addition to intercepting framework operations, \protocol{} can intercept GPU APIs, such as kernel launches, memory copies, and memory allocation/free operations.
To enable it, profilers indicate the domain as \texttt{DLMONITOR\_GPU} to the \texttt{dlmonitor\_callback\_register} function.
Profilers that register callbacks for GPU APIs can capture not only arguments and results about lower-level APIs, but can obtain frames between the low-level APIs and framework operations.
\protocol{} registers callbacks using CUPTI for Nvidia GPUs and RocTracer for AMD GPUs.
To extend \protocol{} for hardware that does not have a vendor-provided callback mechanism, users can define the function signature of the driver function and its in a configuration file.
\protocol{} will register custom callbacks using \texttt{LD\_AUDIT} for all functions recorded in the configuration file.

\paragraph{Call Path Integration}
One key innovation of \protocol{} is its ability to assemble a unified call path that spans from high-level Python code to low-level GPU kernel execution.
At each interception point, if \texttt{dlmonitor\_callpath\_get} is called, \protocol{} retrieves call paths from multiple sources.
The Python call path is obtained using CPython's \texttt{PyFrame}-related APIs.
The ``native'' call path, which includes C/C++ function symbols, is captured using \textit{libunwind}~\cite{libunwind}.
The framework call path is maintained via a shadow stack in each CPU thread.
\protocol{} updates the stack for operators as they are entered and exited, along with their corresponding memory locations.
\texttt{dlmonitor\_callpath\_get} also allows users to choose which specific call path source to integrate or ignore to reduce overhead.

Next, \protocol{} integrates these three call paths into a single comprehensive call path.
It traverses the native call path in a bottom-up direction, matching the address of each frame with the recorded addresses of deep learning operators.
If a match is found, \protocol{} inserts the operator name under the caller frame.
If a frame’s address falls within the \texttt{libpython.so} address space (recorded using \texttt{LD\_AUDIT}), all frames above it are replaced with the Python call path.
If we are at a GPU kernel launch callback, we read parse the function object (e.g., \texttt{CUfunction}) to obtain the kernel name and insert it to the bottom of the call path.

\paragraph{Optimizations}
We have implemented optimizations to provide more insights and reduce the overhead of \protocol{}.

\textit{Forward and backward operator association.}
In PyTorch, when training is enabled, the \texttt{backward} method initiates backward propagation from the leaf operators to the root of the computation graph.
New CPU threads, called \textit{backward threads}, are created for each GPU device to handle this process.
As a result, the native call path of an operator obtained from a backward thread loses the original context of the forward operator, with no Python source code available.
This issue is particularly problematic for GPU kernels without meaningful names, such as the \texttt{elementwise} kernel in PyTorch, as it becomes difficult to trace which source code triggered these calls.
To address this, we record each forward operator’s call path and sequence ID.
In PyTorch, all backward operators associated with a forward operator share the same sequence ID.
Using the sequence ID, the backward thread looks up the forward CPU thread, fetches the Python and framework call path of the corresponding forward operator, and integrates it with the native call path of the backward operator obtained from the backward thread itself.

\textit{Call path caching.}
Unwinding call paths from multiple sources can be costly, especially when GPU APIs are frequently invoked.
To reduce the overhead, we observe that many deep learning operators trigger multiple GPU kernels such that they share the same Python and operator call paths.
Based on the observation, we can reduce redundant call path invocations by caching the Python call path and the deep learning operator in a thread-local variable when an operator is first entered.
Two modes are available: If native call path collection is disabled, we concatenate the shadow call path, GPU API, and GPU kernel function with the cached Python call path.
Otherwise, we retrieve native frames step-by-step using libunwind's \texttt{unw\_step}, building the native call path in a bottom-up direction until we reach the cached deep learning operator.
We then concatenate the cached call path with the native call path.

\subsection{Profiler}
\label{subsec:profiler}
\begin{figure}[tp]
    \centering
    \includegraphics[width=\linewidth]{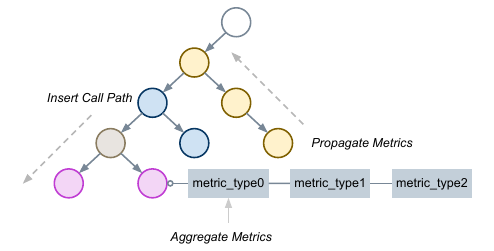}
    \caption{Operations for building a calling context tree and adding performance metrics.}
    \label{fig:calling context}
\end{figure}

In this section, we describe how \tool{}'s profiler collects GPU and CPU metrics and attributes them to a calling context tree as shown in Figure~\ref{fig:calling context}.

\paragraph{Calling Context Tree}
The calling context tree is constructed by inserting call paths obtained from \protocol{} and collapsing frames that refer to the same locations.
For C/C++, GPU API, and GPU kernel frames, two frames are considered the same if they share the same library path and program counter (PC).
For Python frames, they are compared by file path and line number, while framework-based frames are checked by operator names.
Each node in the calling context tree maintains a list of metrics, which are aggregated by sum, minimum, average, and standard deviation for metrics of the same type.
Once a metric has been updated at the bottom of a call path, it is propagated to the root node of the calling context tree, updating the metric along the entire call path.

\paragraph{GPU Metrics}
\tool{} can collect both coarse- and fine-grained metrics on Nvidia and AMD GPUs.
Example coarse-grained metrics include time, parallelism, and shared memory usage, and fine-grained metrics include instruction samples of GPU kernels.
It first registers callbacks for GPU APIs using \protocol{}, then specifies the metrics to profile by calling CUPTI or RocTracer APIs.
At each callback, the profiler emits a unique correlation ID, retrieves the call path, and associates the correlation ID with the call path.
GPU metrics are gathered asynchronously without blocking GPU API calls from the CPU.
When the GPU buffer storing metrics is full, \tool{} flushes the metrics, using the correlation ID to link and aggregate them with the corresponding call path.
Note that if fine-grained metrics, such as instruction samples, are collected, we will extend the call path by inserting the PC of each instruction collected.

\paragraph{CPU Metrics}
\tool{} can profile GPU metrics and CPU metrics in the same run using Linux system calls or CPU measurement substrates.
For example, \tool{} invokes the \texttt{sigaction} system call to registers a signal callback for \texttt{CPU\_TIME} and \texttt{REAL\_TIME} events.
Once a sample is triggered, it will get the current CPU or REAL time, subtract the previous timestamp from it, and use the result as the interval between two samples.
Next, \tool{} will obtain the current call path by calling \texttt{dlmonitor\_callpath\_get} and associate the interval with the call path.
The profiler can also register Linux perf events or invoke PAPI API to obtain metrics from hardware counters.

\subsection{Performance Analyzer}
\label{sec:performance analyzer}

The performance analyzer of \tool{} provides a comprehensive framework for analyzing the profile results and identifying potential performance issues in deep learning workloads.
It initializes the analysis environment by retrieving function symbols from binaries, analyzes control flows, and maps GPU/CPU instructions back to the source code using the DWARF information.
We have designed flexible Python interface to allow performance analysis from the following three key dimensions:
\ecircled{1} \textit{Program Structure Analysis}, which traverses calling contexts and matches call paths given patterns of functions, lines, and metrics; \ecircled{2} \textit{Model Analysis}, which analyzes performance metrics at semantic nodes, such as training, inference, and loss; and \ecircled{3} \textit{Operator Analysis}, which delves into the efficiency of individual operators.
This multi-dimensional approach enables users to gain comprehensive insights, from high-level program behavior down to framework specific behavior. 

\paragraph{Analysis API}
Typically, users instantiate a custom analysis code through the following steps: \textit{call path search}, \textit{metrics analysis}, and \textit{visualization}.
Each analysis starts with the call path search phase. This phase traverses the calling context tree of the profiled application and identifies specific semantic nodes, such as backward/forward computation operations, memory copy operations, loss functions, and evaluation functions, as well as program structure patterns, such as loops and function invocations.
It then applies pattern-matching rules to locate call paths containing these nodes.
In the next, we query the metric data associated with the tree nodes in the matched call paths and apply custom filters to detect potential issues.
Finally, the identified issue nodes are flagged with warning messages, and the detected problems are reported in the GUI.



\paragraph{Example Analyses}
Beyond custom analysis created by users, \tool{} implements a set of example analyses to detect common performance issues using the analysis API.
Below we demonstrate some of the example analyses:
\begin{enumerate}[label=\textbf{\arabic*.}, leftmargin=0pt, align=left, labelwidth=0pt, labelsep=0pt, itemindent=0pt]
\item [\circled{1} \textit{Hotspot Identification}]\mbox{}\\
This analysis identifies the nodes that spend more time than a given threshold and returns their call paths.
\begin{lstlisting}[language=Python]
total_time = call_tree.root.time 
for n in call_tree.kernels:
  if n.time / total_time > hotspot_threshold:
    flag_hotspot(n)
\end{lstlisting}

\item [\circled{2} \textit{Kernel Fusion Analysis}]\mbox{}\\
This analysis detects potential inefficiencies caused by the launch of many small kernels by identifying frames that contain a large number of kernels with short GPU execution times.
\begin{lstlisting}[language=Python]
for n in bfs(call_tree.nodes):
  if n.gpu_time / n.count < gpu_threshold:
    flag_issue(n, "Small GPU kernels")
\end{lstlisting}

\item [\circled{3} \textit{Forward/Backward Operator Analysis}]\mbox{}\\
This analysis identifies deep learning operators whose backward pass takes significantly longer than the forward pass, potentially indicating optimization opportunities because backward phase shouldn't take significantly longer than its forward counterpart.
\begin{lstlisting}[language=Python]
for n in call_tree.operators:
  if n.backward.time / n.forward.time > 2:
    flag_issue(n, "Backward abnormality")
\end{lstlisting}

\item [\circled{4} \textit{Fine-grained Stall Analysis}]\mbox{}\\
This analysis identifies fine-grained stall reasons within hotspot GPU kernels.
Stall reasons for each GPU instruction can be collected using the instruction sampling APIs available on Nvidia and AMD GPUs.
\begin{lstlisting}[language=Python]
hotspots = hotspot_analysis(call_tree)
stalls = []
for n in hotspots:
  for c in n.children:
    if c.stalls > stall_threshold:
      stalls.append(c)
stall_reasons = topk(stalls)
flag_issue(n, "Kernel is mainly stalled by {stall_reasons}")
\end{lstlisting}

\item [\circled{5} \textit{CPU Latency Analysis}]\mbox{}\\
This analysis traverses the calling context tree in the top-down manner to identify frames whose CPU time is significantly higher than GPU time, indicating potential imbalanced workload or synchronization issues.
\begin{lstlisting}[language=Python]
for n in bfs(call_tree.nodes):
  if n.cpu_time / n.gpu_time > cpu_threshold:
    flag_issue(n, "CPU time abnormality")
\end{lstlisting}
\end{enumerate}


\subsection{GUI}
\label{imp_visualization}
\tool{}'s visualization component is a crucial element in presenting profiling results and insights, designed to seamlessly integrate into the developer's workflow, emphasizing on efficiency, responsiveness, and cross-platform compatibility.

The visualization system incorporates the following key features to provide comprehensive insight into program performance:
\begin{itemize}
\item \textit{Flame Graph Visualization.}
We visualize calling context tree in flame graphs~\cite{gregg2016flame} with switchable top-down and bottom-up views, enabling analysis for different purposes.
The top-down view provides a direct representation of the calling context tree, while the bottom-up view aggregates individual metrics at the same node across different call paths.
In both views, we highlight hotspot call paths using the hotspot analysis described in the previous section.
\item \textit{Interactive Performance Analysis.}
We use a color-coded system to highlight issues reported by the performance analyzer, ensuring that users are alerted to critical bottlenecks.
Users can click on a highlighted frame to access all related metrics and view the corresponding source code on demand, significantly accelerating the optimization process.
\end{itemize}

Our GUI is developed using the VSCode API, enabling its use not only in VSCode but also in other IDEs that implement the VS Code protocol, such as VSCodium~\cite{VSCodium} and Eclipse Theia~\cite{EclipseTheia}. 
The implementation of our GUI consists of two key modules, \textit{WebView-based Visualization Interface} and \textit{IDE Interaction}.
This WebView-based Visualization Interface combines HTML-based text rendering for clear and formatted textual information with WebGL-powered graphical rendering for high-performance visual representations of performance data.
The IDE Interaction module manages the connection between the visualization interface and the IDE's core functionalities.
When activated, the backend translates visualization events (such as clicking on a function hotspot) into precise editor actions, such as opening a file, navigating to the corresponding line, and highlighting relevant sections, streamlining the process of identifying and addressing performance issues.

\section{Evaluation}
\label{evaluation}

\paragraph{Platforms.} We evaluated \tool's overhead on two platforms equipped with different GPUs shown in Table~\ref{tab:evaluation_platforms}.

\begin{table*}[t]
    \footnotesize
    \centering
    \caption{Evaluation Platforms}
    \begin{tabular}{c c c c c c}
    \toprule
        \makecell{\textbf{Platform}}
        & \makecell{\textbf{CPU}}
        & \makecell{\textbf{Memory}}
        & \makecell{\textbf{GPU}}
        & \makecell{\textbf{GPU Memory}}
        & \makecell{\textbf{GPU Specifications}} \\
    \midrule
        \makecell{\textbf{Nvidia}} & \makecell{AMD EPYC 7543} & \makecell{256 GB} &
        \makecell{A100 SXM} & \makecell{80 GB} &
        \makecell{108 SMs, 156 TF32 TFLOP/s, 2TB/s Bandwidth} \\
    \lightmidrule
        \makecell{\textbf{AMD}} & \makecell{AMD EPYC 7543} & \makecell{2048 GB} &
        \makecell{MI250} & \makecell{64 GB} &
        \makecell{208 Compute Units, 362.1 FP16 TFLOP/s, 3.2 TB/s Bandwidth} \\
    \bottomrule
    \end{tabular}
    \label{tab:evaluation_platforms}
\end{table*}

\begin{table*}[t]
    \centering
    \footnotesize
    \caption{Case Studies Summary}
    \begin{tabular}{c c c l l c}
    \toprule
    \makecell{\textbf{Deep Learning Model}}
    & \makecell{\textbf{Dataset}}
    & \makecell{\textbf{Platform}}
    & \makecell{\textbf{Analysis Client}}
    & \makecell{\textbf{Optimization Method}}
    & \makecell{\textbf{Speedup}} \\
    \midrule
     DLRM-small & Criteo 1TB & Nvidia & \makecell{\circled{3} Forward/Backward Operator Analysis} & \makecell{replace \texttt{aten::index} \\ with \texttt{aten::index\_select}} & 1.66$\times$ \\
     \lightmidrule
     GNN & \makecell{OGBG-MOLPCBA} & Nvidia &  \makecell{\circled{3} Forward/Backward Operator Analysis} & \makecell{replace \texttt{aten::index} \\ with \texttt{aten::index\_select}}& 1.07$\times$ \\
     \lightmidrule
     UNet & \makecell{fastMRI} & Nvidia & \makecell[l]{\circled{1} Hotspot Identification} & \makecell{Avoid \texttt{channel\_first} to \\ \texttt{channel\_last} conversion} & 1.28$\times$ \\
     \lightmidrule
     UNet & \makecell{fastMRI} & Nvidia & \makecell[l]{\circled{5} CPU Latency Analysis} & \makecell{Match \texttt{worker\_num} \\ with \texttt{\#CPU} cores} & 1.15$\times$ \\
     \lightmidrule
     Transformer-Big & \makecell{WMT} & Nvidia & \makecell[l]{\circled{2} Kernel Fusion Analysis} & \makecell{Fuse small kernels \\ using \texttt{torch.compile}} & 1.06$\times$ \\
     \lightmidrule
     Llama3 & \makecell{Sample Prompt} & Nvidia & \makecell[l]{\circled{4} Fine-grained Stall Analysis} &  \makecell{Use fast data type\\ conversion instructions} & N/A \\
     \lightmidrule
     UNet & \makecell{fastMRI} & AMD \& Nvidia & \makecell[l]{\circled{1} Hotspot Identification} & \makecell{Adjust number of threads\\ per CTA} & N/A \\
     \lightmidrule
     \makecell{DLRM-small \\ GNN \\ UNet \\ ResNet} & \makecell{Criteo 1TB \\ OGBG-MOLPCBA \\ fastMRI \\ ImageNet} & \makecell{Nvidia-JAX \\ Nvidia-PyTorch} & \makecell[l]{\circled{2} Kernel Fusion Analysis} & \makecell{Fuse small kernels\\
using \texttt{torch.compile}} & N/A \\
    \bottomrule
    \end{tabular}
    \label{tab:case_studies_summary}
\end{table*}

\paragraph{Workloads.} We used \tool to profile the MLCommons Algorithm Efficiency benchmark~\cite{Dahl2023AlgoPerf}, implemented in both PyTorch~\cite{ansel2024pytorch} and JAX~\cite{jax2018github}. 
We evaluated the eager mode of PyTorch and the JIT mode of JAX.
We run each model for 100 iterations using different profiling tools.

The following workloads and datasets were used.
\begin{itemize}
  \item Conformer~\cite{gulati2020conformer} with the LibriSpeech~\cite{panayotov2015librispeech} dataset.
  \item DLRM-small~\cite{naumov2019deep} with the Criteo 1TB~\cite{criteo2014} dataset.
  \item U-Net~\cite{ronneberger2015u} with the fastMRI~\cite{zbontar2018fastmri} dataset.
  \item GNN~\cite{battaglia2018relational} with the OGBG-MOLPCBA~\cite{hu2020open} dataset.
  \item ResNet~\cite{he2016deep} with the ImageNet~\cite{deng2009imagenet} dataset.
  \item Vision Transformer~\cite{dosovitskiy2020image} with the ImageNet dataset.
  \item Transformer-Big~\cite{vaswani2017attention} with the~WMT\cite{bojar2017findings} dataset.
  \item Llama 3~\cite{dubey2024llama} inference with a sample prompt from huggingface official example.
  \item Gemma~\cite{team2024gemma} with the same prompt as Llama 3.
  \item nanoGPT~\cite{Karpathy2022} with the same prompt as Llama 3.
\end{itemize}

\begin{figure*}
    \centering
    \begin{subfigure}[h]{\linewidth}
        \includegraphics[width=\linewidth]{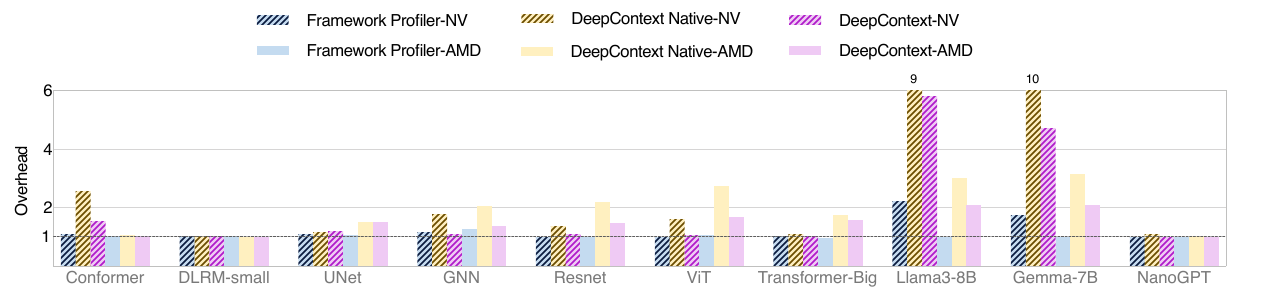}
        \caption{Time overhead of PyTorch workloads using different profilers on Nvidia and AMD GPUs.}
    \end{subfigure}
    \begin{subfigure}[h]{\linewidth}
        \includegraphics[width=\linewidth]{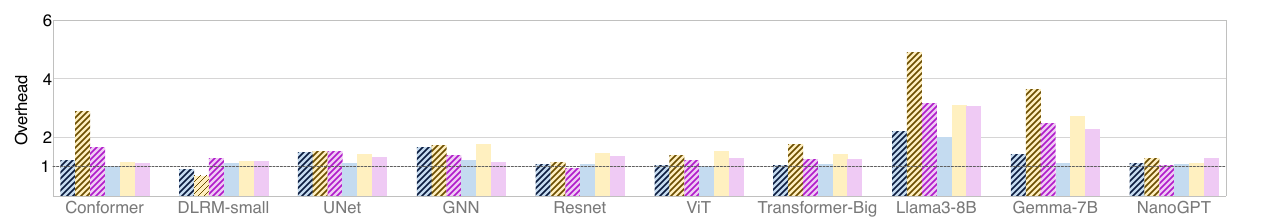}
        \caption{Time overhead of JAX workloads using different profilers on Nvidia and AMD GPUs.}
    \end{subfigure}
    \begin{subfigure}[h]{\linewidth}
        \includegraphics[width=\linewidth]{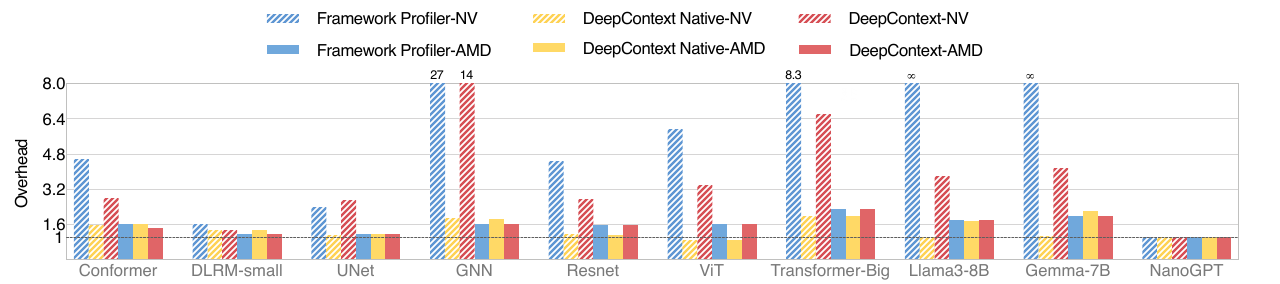}
        \caption{Memory overhead of PyTorch workloads using different profilers on Nvidia and AMD GPUs.}
    \end{subfigure}
    \begin{subfigure}[h]{\linewidth}
        \includegraphics[width=\linewidth]{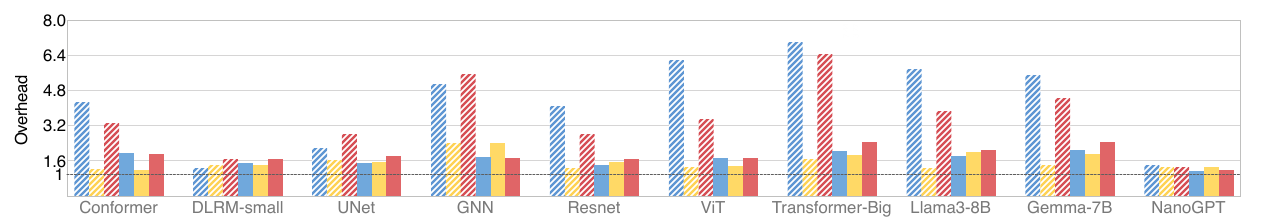}
        \caption{Memory overhead of JAX workloads using different profilers on Nvidia and AMD GPUs.}
    \end{subfigure}
    \caption{Comparison of the time and memory overheads of various workloads using \tool{} with the PyTorch profiler and the JAX profiler.}
    \label{fig:overhead_graph_total}
\end{figure*}

\paragraph{Results} We measured the end-to-end running time of each workload under three circumstances: without profiler enabled, \tool with Python and framework call paths obtained from \protocol, and \tool with Python, deep learing framework, and native C/C++ call paths.
Then we divide the running time of \tool enabled by the running time without \tool enabled to calculate the overhead, as shown in Figure~\ref{fig:overhead_graph_total}.

The median running time overhead of \tool{} is 1.12$\times$ and 1.50$\times$ for PyTorch on Nvidia and AMD GPUs, respectively. 
For JAX, its median overhead is 1.33$\times$ and 1.28$\times$ on Nvidia and AMD GPUs, respectively. 
When the C/C++ native call path is not collected, we observed median overheads of 1.50$\times$ and 1.90$\times$ for PyTorch, and 1.60$\times$ and 1.46$\times$ for JAX, on Nvidia and AMD GPUs, respectively.
The overhead with native call path is higher than the variant without the native call path due to the additional overhead in unwinding C/C++ call paths and concatenating them with Python and framework call paths.

In comparison, PyTorch profiler incurs a median overhead of 1.06$\times$ and 1.01$\times$ on Nvidia and AMD GPUs, respectively.
JAX profiler incurs a median overhead of 1.17$\times$ and 1.10$\times$ on Nvidia and AMD GPUs, respectively.
Without native call path collection, the overhead of \tool{} is comparable to that of framework profilers.
We do observe a much higher time overhead from profiling Llama3 and Gemma-7B using PyTorch.
The overhead is caused by two factors: our frame unification system, which identifies the same file path and line number, and our metrics aggregation and propagation mechanism along the call paths, introduces additional overhead.
These two factors are especially significant with such workloads launch many small kernels.

The median memory overhead of \tool{} is 1.00$\times$-2.44$\times$, compared with that of 1.29$\times$-27.28$\times$ and 1.27$\times$-6.98$\times$ of PyTorch and JAX profilers, respectively.
Note that the memory overhead of the framework profilers will increase with the increase in the number of iterations.
Also, the PyTorch profiler encountered out-of-memory issues when exporting the profiling database to disk, failing to provide any insights for optimization.
\tool{} incurs significant lower memory overhead compared to these tools because it aggregates metrics at runtime and thus is more feasible for long-running workloads.
 


\section{Case Studies}
\label{sec:case studies}

In this section, we describe seven use cases conducted using optimization insights obtained from \tool, summarized in Table~\ref{tab:case_studies_summary}.
All case studies were conducted by a graduate student who has experience using PyTorch and JAX but no experience with the low-level code of these frameworks.
Our observations and optimizations have been verified by an experienced Systems researcher with extensive experience in deep learning systems and GPUs.

\subsection{Forward/backward Operator Analysis}

\begin{figure}[h]
    \centering
    \includegraphics[width=\linewidth]{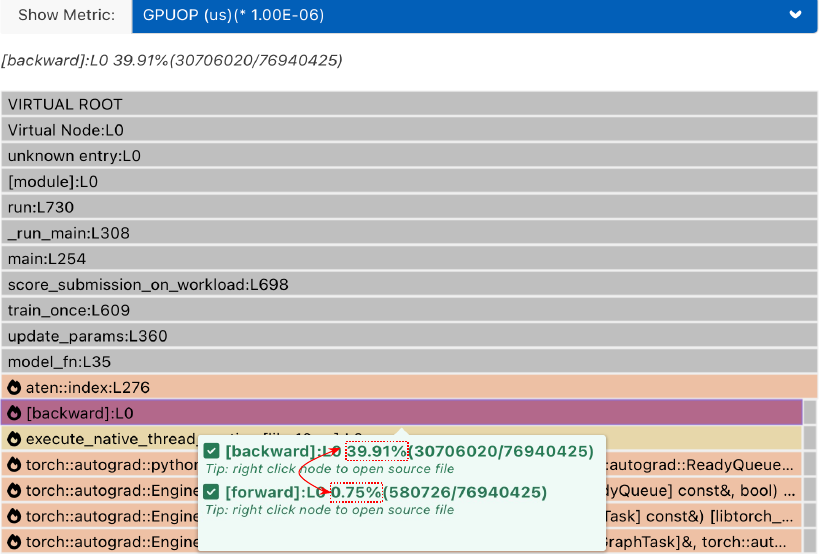}
    \caption{Forward-Backward association view of the DLRM-small workload.}
    \label{fig:top_down_view}
\end{figure}

We profiled the DLRM-small workload using the Criteo 1TB dataset on the A100 platform.
In \tool's bottom up view, we noticed that the hotspot is on the \\
\texttt{indexing\_backward\_kernel} kernel (30.5s), which takes 39.6\% of the total GPU kernel time.
Using the \tool's framework call path, which associates the forward call path with the corresponding backward kernels, we found that this GPU kernel is triggered by the backward computation of \texttt{aten::index} called by \texttt{embedding\_table[idx\_lookup]}, as illustrated in Figure~\ref{fig:top_down_view}.

It should be noted that while the backward computations of \texttt{aten::index} take 39.9\% time, the forward computation takes only 0.8\% time.
This discrepancy is caused by the deterministic nature~\cite{Gross2020} of \texttt{aten::index}, which serializes GPU threads accessing the same memory location and is unnecessary in this workload if determinism is not required.
To optimize the code, we substituted \texttt{aten::index} with a non-deterministic operator \texttt{aten::index\_select}, which uses atomic operations in the backward phase to avoid serialization and \textbf{reduced the total GPU time from 73.2s to 44.0s}.
We have also observed the same problem in the GNN workload; applying the same optimization \textbf{reduced the total GPU time from 3.97s to 3.71s}.

\subsection{Hotspot Identification with Call Path}

When profiling U-Net using the fastMRI dataset on the A100 platform, we observed that the \texttt{cudnn::nchwToNhwcKernel} kernel takes 15.4\% of the GPU time.
\begin{figure}[h]
    \centering
    \includegraphics[width=\linewidth]{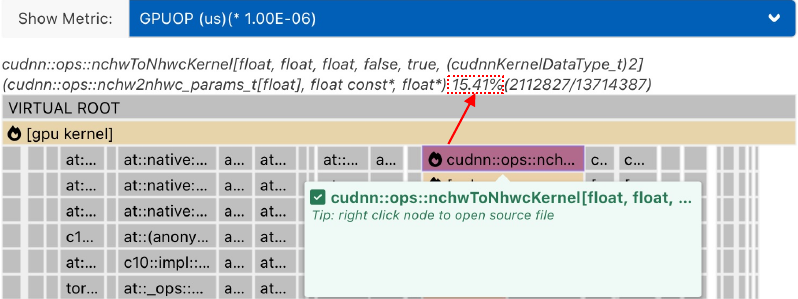}
    \caption{The bottom-up view of U-Net.}
    \label{fig:unet_bottom_up_view}
\end{figure}
Using \tool's framework and Python call paths, we identified every PyTorch operator that invokes the conversion.
In addition, with the help of native call paths, we also identified that the input tensor's memory format is converted from PyTorch's default \texttt{channels\_first} layout ~\cite{pytorch_memory_format_tutorial} to the \texttt{channels\_last} layout---a layout more efficient layout for CUDNN---and then reverted back to \texttt{channels\_first} after the computations, introducing excessive overhead.
To address memory format conversion issues, we optimized the code by storing input tensors with \texttt{channels\_last} layout before computations, and refactored LayerNorm and InstanceNorm layers to store weights in the \texttt{channels\_last} layout to avoid conversion. This optimization reduced the \textbf{end-to-end time of 100 iterations from 54s to 42s}.

\subsection{Kernel Fusion Analysis}

Using \tool, we profiled the Transformer-Big workload on the A100 platform.
\tool can gather multiple metrics in a single run, such as the number of invocations, the number of warps and blocks, as well as the number of shared memory and registers, in addition to the GPU time. 
These metrics are attributed to the corresponding frames in the call paths to assist performance analysis.

\begin{figure}
    \centering
    \includegraphics[width=\linewidth]{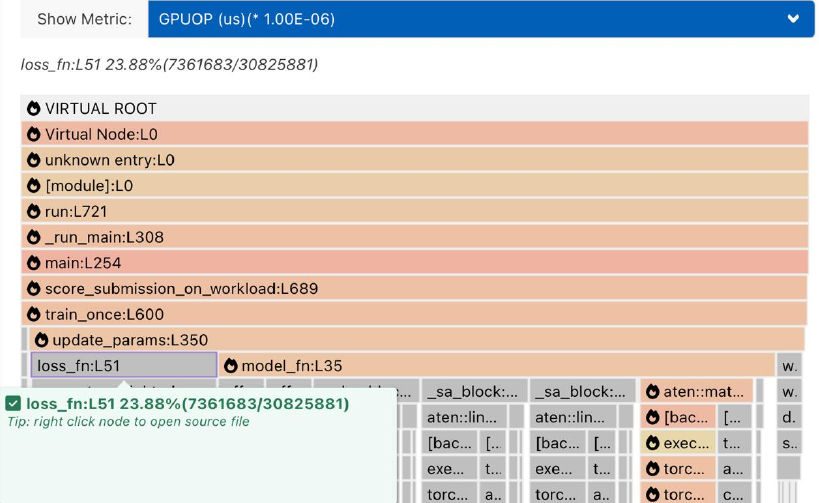}
    \caption{Top-down view of Transformer-Big}
    \label{fig:wmt_loss_fn}
\end{figure}

For instance, from the top-down view in Figure \ref{fig:wmt_loss_fn}, we observed that \texttt{loss\_fn} takes 7.36s, which is 23.9\% of total time.
Under this frame, there are three different kernels invoked: including \texttt{softmax}, \texttt{copy}, and \texttt{nll\_loss}, each with the same number of invocations.
The kernel fusion analysis suggests an opportunity for optimization by combining small kernels to reduce overall time.
Further analysis expanding the call paths also reveals that the \texttt{softmax} kernel has relatively low register usage, which implies that fusing this kernel will not cause significant register overhead.
Based on the suggestion and observation, we manually fused these small kernels into a single and more efficient kernel. 
After optimization, the \textbf{total GPU time is decreased from 30.5s to 23.9s.}

\subsection{CPU Latency Analysis}

We enabled both CPU and GPU metrics to profile the U-Net workload.
The CPU latency analysis highlighted that the call path to the \texttt{data\_selection} function takes 69\% of the CPU time with 16 threads running concurrently, while the GPU time of the same frame is only 1.3 seconds.
Further investigation shows that the first iteration of loading data from the disk to the memory takes 10 seconds, and the GPU remains idle.
By expanding the call paths of \texttt{data\_selection}, we noted that an inefficient setting of parallel threads has been invoked. Our allocated node only has 6 physical CPU cores, but the data loader is hard coded with 16 threads to load the data, causing significantly scheduling overhead.
After we reduced the thread number to 8, we reduced \textbf{the end-to-end time of 100 iterations by 7s, from 54s to 47s}.


\subsection{AMD \textit{vs} Nvidia}

\begin{figure}[h]
    \begin{subfigure}[hb]{\linewidth}
        \includegraphics[width=\linewidth]{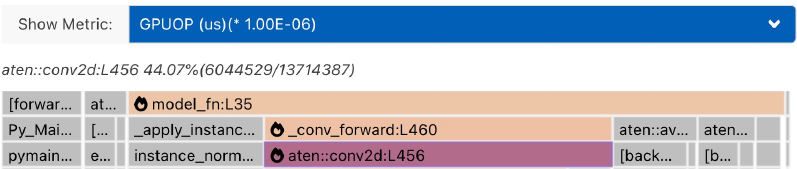}
        \caption{Flame Graph of Nvidia GPU. Hotspot is on operator \texttt{aten::conv2d}, which is as expected.}
        \label{fig:nv_flame_graph}
    \end{subfigure}
    \centering
    \begin{subfigure}[hb]{\linewidth}
        \includegraphics[width=\linewidth]{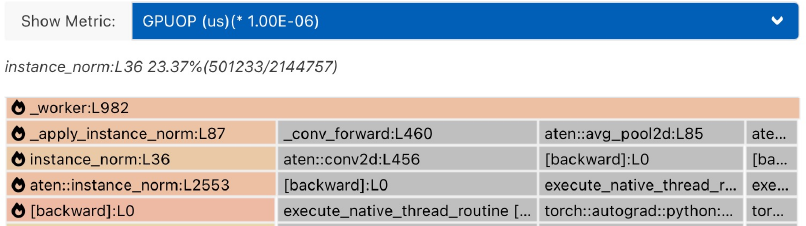}
        \caption{Flame Graph of AMD GPU. Hotspot is on operator \texttt{aten::instance\_norm}, which is abnormal.}
        \label{fig:amd_flame_graph}
    \end{subfigure}
    \caption{AMD v.s. Nvidia}
    \label{fig:amd_vs_nv}
\end{figure}

\sloppy
We profiled the U-Net workload on both AMD and Nvidia GPUs. From the top-down view shown in Figure \ref{fig:amd_vs_nv}, we can see differences between these two platforms.
Figure \ref{fig:nv_flame_graph} shows that on Nvidia GPUs, the performance hotspot is on convolution operator \texttt{aten::conv2d}; On the other hand, Figure \ref{fig:amd_flame_graph} shows that on AMD GPUs, the performance hotspot lies on instance norm operator \texttt{aten::instance\_norm}.
In order to find out the cause of performance degradation on AMD GPU, we checked the low level call paths of \texttt{aten::instance\_norm} provided by \tool and found that the implementation of \texttt{aten::instance\_norm} for AMD GPUs provided by PyTorch reused the same kernel template---\texttt{batch\_norm\_backward\_cuda\_template}---as that for Nvidia GPUS~\cite{pytorch2024normalization}, with the same number of threads per CTA.
Since AMD and Nvidia GPUs have different architectures—AMD GPUs have a warp size of 64, while Nvidia GPUs have a warp size of 32—AMD kernel has fewer CTAs than Nvidia GPUs, resulting in lower parallelism.
To optimize the kernel on AMD GPUs, we can adjust the number of threads to best fit the different architectures.

\subsection{JAX \textit{vs} PyTorch}
We compared the performance of JAX and PyTorch across four datasets/models: DLRM-small, U-Net, GNN, and ResNet.
Our results show that JAX significantly outperforms PyTorch in all tasks, achieving performance improvements exceeding 50\%. By comparing the number of kernel operations, we observed that the JAX version consistently requires fewer operations than its PyTorch counterpart. This substantial performance gap is primarily attributed to the advantages of JAX's XLA compiler, which effectively fuses operators to reduce redundant memory access and overlap compute and memory instructions.
To narrow this performance gap, we can use \texttt{torch.compile} to optimize some workloads, but we observed that it cannot be enabled for all cases and will incur long autotuning overhead using the \texttt{max-autotune} mode.

\subsection{Fine-grained Stall Analysis}
We profiled the Llama3 workload running low-precision including \texttt{float16} and \texttt{float8} using fine-grained instruction sampling.
On both AMD and Nvidia GPUs, we have identified time spent on the data conversion operators (i.e., \texttt{torch.to}) in the \texttt{LlamaRMSNorm} module~\cite{huggingface2024llama}.
The fine-grained stall analysis identifies non-trivial constant memory misses due to the loading of constants for each CTA.
Since the input is small, there is a relatively high overhead in reading the constant memory compared to loading the data itself.
Additionally, we observed math dependency-related stalls caused by non-vectorized data conversion instructions from/to \texttt{float32}. 
To optimize the kernel, we can (1) ensure that each block loads the minimum number of bytes required to use vectorized data conversion instructions, and (2) fuse the conversion operator with other operators to ensure that the constant memory overhead is minimized.
\section{Conclusions}
\label{sec:conclusions}

\tool{} addresses a critical gap in performance profiling for deep learning workloads in heterogeneous computing environments, where the interaction between CPUs, GPUs, and deep learning frameworks is inherently complex.
\tool{} fulfills this need by providing a multi-level, automated analysis that bridges the different layers of the software and hardware stack.
Our detailed case studies and evaluations show that \tool{} improves the ability to identify and resolve performance bottlenecks in deep learning workflows.

Currently, the primary limitation of \tool{} lies in the overhead introduced when profiling workloads with small kernels, where the cost of unwinding the call path becomes significant.
To address this, we plan to explore the call path caching techniques~\cite{zhou2022low}.
In the future, we also aim to extend \tool{} to support PyTorch workloads that use \texttt{torch.compile}, applying similar profiling methods for JAX to capture the call paths of deep learning operators.


\bibliographystyle{plain}
\bibliography{references}

\end{document}